\documentclass[preprint2]{aastex}
\usepackage{graphicx}
\usepackage{natbib}

\shorttitle{Infrared polarimetry of exoplanets}
\shortauthors{de Kok et al.}

\begin{document}

\title{Characterizing exoplanetary atmospheres through infrared polarimetry}

\author{R.J.~de Kok, D.M.~Stam, and T.~Karalidi}
\affil{SRON Netherlands Institute for Space Research, Sorbonnelaan 2, 3584 CA Utrecht, the Netherlands}
\email{R.J.de.Kok@sron.nl}

\begin{abstract}
Planets can emit polarized thermal radiation, just like brown dwarfs. We present calculated
thermal polarization signals from hot exoplanets, using an advanced radiative transfer code
that fully includes all orders of scattering by gaseous molecules and cloud particles. The code spatially resolves the disk of the planet, allowing simulations for horizontally inhomogeneous planets.
Our results show that the degree of linear polarization, $P$, of an exoplanet's thermal radiation 
is expected to be highest near the planet's limb and that this $P$ depends on the temperature and 
its gradient, the scattering properties and the distribution of the cloud particles. 
Integrated over the disk of a spherically symmetric planet, $P$ of the thermal radiation
equals zero. However, for planets that appear spherically asymmetric, e.g. due to flattening, 
cloud bands or spots in their atmosphere, differences in their day and night sides, and/or
obscuring rings, $P$ is often larger than 0.1\%, in favorable cases even reaching several
percent at near-infrared wavelengths.
Detection of thermal polarization signals can give access to planetary parameters that are 
otherwise hard to obtain: it immediately confirms the presence of clouds, and
$P$ can then constrain atmospheric inhomogeneities and the flattening due to the planet's rotation rate.
For zonally symmetric planets, the angle of polarization will yield the components of the
planet's spin axis normal to the line-of-sight. Finally, our simulations show that $P$ is generally more sensitive to variability in 
a cloudy planet's atmosphere than the thermal flux is, and could hence better reveal certain dynamical processes.
\end{abstract}


\section{Introduction}

Studying the thermal emission of exoplanets has only recently become possible with the 
direct detections of young giant planets in wide orbits around their star 
\citep[e.g.][]{mar08,laf08} and secondary eclipse detections of transiting exoplanets 
in tight orbits \citep[see][]{dem09}. These hot planets emit most of their radiation at 
near-infrared wavelengths and flux measurements at different wavelengths are used to 
constrain properties of their atmospheres. 
However, little attention has been given to the information contained within the polarization signals of this planetary thermal radiation.

Incident starlight will usually get polarized when it is 
scattered by gases and particles in a planetary atmosphere. When integrated over
the planetary disk, the reflected starlight can yield a significant net degree
of polarization \citep[e.g.][]{sea00,sta04}.
Also polarization in stellar atmospheres is well-known \citep[e.g.][]{har70} and  a dozen polarized brown dwarfs have been identified \citep[e.g.][]{men02}. Thermal radiation that is emitted by a body can get polarized upon scattering.
For a net polarized thermal signal of a spatially unresolved body such as
a brown dwarf or an exoplanet to be observable, not only scattering particles are required, but also an 
asymmetry in the body's disk. Otherwise, the polarized thermal signals from different parts
of the disk will cancel each other completely. With an asymmetric disk, the cancellation
will be incomplete and a net polarization signal remains.

For the polarized brown dwarfs, a plausible source of asymmetry is the flattening
of the body due to its rotation, as advocated by \citet{sen10}. Very recently, these authors have extended their work on polarization of flattened objects to planets as well \citep{mar11}.
In a planetary atmosphere, sources of asymmetry could arise from horizontal
inhomogeneities in temperature or cloud thickness.
For instance, strong zonal winds can cause banded structures, as found on all giant 
planets in our solar system. Furthermore, rings that obscure or shadow part of the
planetary disk would cause asymmetries.  

Here, we will use simulated signals of horizontally inhomogeneous planets to 
present processes that polarize thermal planetary radiation, to explore 
parameters that determine the strength of these thermal polarization signals, and
to discuss the value of infrared polarimetry for the characterization of exoplanet atmospheres.


\section{Our numerical model}

Light is fully described by a flux vector ${\bf F}= (F,Q,U,V)$, with $F$ the total flux, 
$Q$ and $U$ the linearly polarized fluxes, and $V$ the circularly polarized flux \citep{han74}. 
Parameters $Q$ and $U$ are defined with respect to a given reference plane. For atmospheric calculations we have a local reference plane with axes parallel and normal to the local planetary horizon.
In the following, we will neglect $V$, and express the degree of polarization, $P$, using
\begin{equation}
   P= \frac{\sqrt{Q^2+U^2}}{F}
\end{equation}
The angle of polarization, $\chi$, with respect to the reference plane is defined as \citep[see][]{han74}
\begin{equation}
\tan 2 \chi = U/Q.
\end{equation}

We calculate the radiative transfer in a locally plane-parallel, vertically inhomogeneous planetary 
atmosphere for a range of emission angles using a doubling-adding method \citep{wau94}, 
which fully includes all orders of scattering and polarization. 
Each atmosphere consists of 40 layers, equally spaced in log pressure, with pressures $p$ between 10$^{-6}$~bar (top) and 5~bar (bottom).
In order to make comparisons between different model atmospheres we use ad hoc temperature 
profiles, assuming hydrostatic equilibrium, which are uniform in temperature $T$ for $p < 10^{-3}$ bar and $p > 1$~bar. In between these pressures the near-infrared emission originates and we assume $dT/d\ln{p}$ is constant there, pivoting around $T= 1500$~K at $p= 33$~mbar for different temperature profiles.
We perform our calculations at $\lambda_1= 1.05$~$\mu$m, which is in the continuum ($Y$-band), 
and at $\lambda_2=1.11$~$\mu$m, in a water vapor absorption band. 
The water absorption optical depths are calculated using the HITEMP 2010 database
\citep{rot10}, assuming Voigt line shapes and a volume mixing ratio of 5$\cdot$10$^{-4}$. 

At these infrared wavelengths, the scattering optical thickness of the gas molecules is negligible, but radiation can be scattered by larger cloud/dust particles.
We calculate the scattering matrices of the particles in our model atmospheres
using Mie-scattering as described by \citet{der84} (thus assuming spherical particles).
The particles that we use are small compared to the wavelength and have unity single
scattering albedo. The scattering of these particles can be described as Rayleigh-scattering,
with a nearly isotropic flux phase function and a bell-shaped polarization phase function
reaching almost 100\% polarization at a single scattering angle of 90$^\circ$.
The particles are assumed to be in one of the atmospheric layers and 
we assume identical cloud optical thicknesses at $\lambda_1$ and $\lambda_2$, which only have a difference in wavelength of 0.05 $\mu$m.
We also performed calculations using larger, non-Rayleigh-like, particles, which will be discussed in 
Section 3.4. 

\section{Spatially resolved polarimetry}

To understand disk-integrated polarization signals of exoplanets better, we first discuss
polarization signals that are spatially resolved across the disk. 

Generally, singly scattered radiation is unpolarized for scattering angles equal or close to 
0$^\circ$ \citep[assuming unpolarized incoming radiation,][]{han74}. Hence, thermal radiation
emitted by low atmospheric layers that is scattered upward by particles in upper layers
will have a very low degree of polarization. 
Radiation that is emitted in a high altitude layer that itself contains scattering 
particles can also be scattered in the upward direction over a 90$^\circ$ angle and be strongly 
polarized, in particular when the scattering is Rayleigh-like. However, because the 
scattering particles in the upper layer will receive emitted radiation from all directions equally, the net polarization of the scattered radiation that emerges from the layer will equal zero. Hence, at the center of a horizontally homogeneous planetary disk, $P$ will usually be low, regardless of the temperature profile.

At the limb of a planetary disk, thermal radiation that is emitted by low atmospheric 
layers will be scattered by particles in upper layers towards the observer over
angles near 90$^\circ$, yielding high values of $P$. Radiation that is emitted
in high altitude layers that contain scattering particles themselves will yield relatively low values of $P$, because of the large range of scattering angles. $P$ will not be zero mainly because of contributions of radiation that has been scattered twice or more.

The degree of polarization at the limb thus depends strongly on the temperature structure of the atmosphere. This is related to the limb darkening effect, which is illustrated in Figure~1. For a given optical path length, radiation arriving at an angle towards the scatterer will come from a higher layer in the atmosphere than the light coming from directly below. If there is a temperature gradient, that difference in altitude will correspond to a difference in temperature and hence a difference in thermal flux. If the temperature decreases with 
decreasing pressure, most thermal radiation emerging from the limb of the planet
will have been emitted in the lower atmospheric layers, and will have been scattered at high altitudes 
towards the observer at angles close to 90$^\circ$, where $P$ of Rayleigh scattered
radiation is highest. If the temperature increases with decreasing pressure, 
most of the thermal radiation that is observed at the planet's limb will have been
emitted and scattered in the upper layers. In this case, the large range of 
scattering angles will yield relatively low values of $P$ (see Figure~1). 
If the temperature gradient equals zero, the radiation field around the scatterers
is more symmetric, even at the limb, and the net $P$ of radiation emitted
towards the observer will be close to zero. 

\subsection{Dependence on temperature gradient}

Figure~2a shows the normalized polarized flux $-Q/F$ for model atmospheres
with different temperature profiles and a high altitude cloud layer with an optical thickness 
of 0.1 at $\lambda_1$ and $\lambda_2$ for small and large emission angles.
In these one-dimensional model calculations, $U=0$, and hence  $|Q/F|=P$. A positive (negative) $-Q/F$ indicates that the radiation is polarized parallel
(perpendicularly) with respect to the local horizon.
The curves in Figure~2a show the strong influence of the temperature gradient on $P$ and that a
positive temperature gradient changes the polarization angle $\chi$ by 90$^\circ$. 

It is also clear that for positive temperature gradients, $P$ is highest 
for $\lambda_2$, the wavelength in the 
water vapor absorption band. This can again be understood in terms of the effect shown in Figure~1. In the continuum, the radiation that arrives at the scattering particle originates deeper in the atmosphere than in the water band. Because of our constant temperature gradient, the temperature difference between radiation coming towards the particle from directly below and that coming at an angle will be roughly similar for both wavelengths. However, such a temperature difference corresponds to a relatively larger difference in flux for lower temperatures, meaning that radiation arriving from the colder atmospheric layers probed in the absorption band will have relatively more contributions from directly below and hence will give a larger $P$ at the limb. For negative temperature gradients, $-Q/F$ is negative, which shows that here the emerging signal is dominated by radiation arriving at an angle (see Figure 1b) that is scattered multiple times. In the continuum, there is now relatively less radiation arriving from directly below, giving less positive contributions to the net $-Q/F$ value. Hence, $P$ is higher in the continuum than in the absorption band for atmospheres with a negative temparature gradient and high clouds.

\subsection{Dependence on cloud optical thickness}

An optically very thin cloud
won't affect the outgoing radiation much as the cloud is almost transparent to the outward-going radiation. Hence, not much radiation is scattered and $P$ will be low. With increasing cloud optical 
thickness, $P$ will increase, because more radiation is scattered as the transmission of the clouds decreases.
On the other hand, the contribution of multiple scattered radiation will also increase with increasing cloud (scattering) optical thickness. This is because the free path length of the radiation can become less than the distance between scattering particles. Multiple scattered radiation arriving at the observer will have been scattered at a range of angles. At the limb, this means that the observed radiation has not only been scattered over angles that produce a high $P$, but it will also have been scattered over angles that produce low $P$.
 For small emission angles, such as observed near the center of the planetary disk, multiple scattering can actually
increase $P$, because it adds radiation that has been scattered at angles where $P$ is large.   For very large 
optical thicknesses, virtually all the thermal radiation will originate from the cloud and be multiple scattered within the cloud itself and $P$ will be relatively low.
These various effects of the cloud optical thickness can be seen in Figure~2b: first $P$ increases with cloud optical thickness as more radiation is scattered, then $P$ decreases as multiple scattering becomes important, and finally $P$ stays roughly constant at large cloud optical thicknesses as virtually all radiation originates from multiple scattering within the cloud itself. 

\subsection{Dependence on cloud top height}

In Figure~2, the cloud is located high in the atmosphere. Lowering the cloud does not 
change $P$ much until the gas absorption optical depth is comparable to the cloud 
scattering optical thickness. For optically thin clouds, $P$ gradually vanishes as the cloud descents and gaseous absorption starts dominating over scattering. 
For thicker clouds, $P$ at the limb will first rise slightly before 
vanishing as multiple scattering effects are reduced. 
Because gas optical thicknesses depend on wavelength, cloud top heights can have large effects
on $P$'s spectral behavior.
For instance, in gaseous absorption bands, $P$ can be zero when clouds are below the altitudes 
where gaseous absorption is high, whereas it can be non-zero at continuum wavelengths. Hence, a case with low clouds can leave peaks in the polarization spectrum at certain wavelengths, whereas in a case with high clouds these wavelengths represent a local minimum in the polarization spectrum.

\subsection{Dependence on scattering properties}

In the previous calculations, we assumed particles that were small relative to the wavelength (i.e.~they are Rayleigh scattering). The polarization signal
of a planet will, however, depend on the particles' scattering properties.  With increasing particle size, the maximum value of the polarization phase functions 
usually decreases and polarization direction changes occur \citep[e.g.][]{sta04}. 
Calculations with Saturn-like particles, whose scattering properties are derived by \citet{tom84}, and 1-$\mu$m spherical Mg$_2$SiO$_4$ 
particles show that $P$ at the limb will decrease to less than 10\% and that the sign of $Q$ will be opposite to that shown in Figure~2. In both of these cases the single scattering albedo of the particles is close to unity.

\subsection{Three-dimensional effects}

The above calculations were performed using one-dimensional models, which are applicable to atmospheres 
that are locally horizontally homogeneous. However, large local variations in e.g.~temperature or
composition can give rise to adjacency effects that affect a planet's polarization signal.
For instance, a hot spot in an atmosphere could be surrounded by a ring of high polarization
resulting from radiation that is emitted by the hot spot, which is subsequently scattered by 
the colder atmosphere around it. Such three-dimensional effects can be the subject for later study.

\section{Disk-integrated polarimetry}

To simulate exoplanet signals, we integrate the spatially resolved fluxes 
across the planetary disk, using a grid with a 2$^\circ$ resolution in latitude and longitude.
The reference plane for the polarized fluxes $Q$ and $U$ 
depends on the location on the planet \citep{wau94}. Before integrating these fluxes, we thus
have to rotate the local flux vectors to the planet's global reference plane \citep{hov83,sta06}, which we align with the planet's spin axis.
For spherically symmetric planets, the rotation and subsequent integration yields a complete 
cancellation of both $Q$ and $U$ of the planetary thermal radiation (see Figure~3).

We have performed calculations for five different asymmetric planets, all illustrated in Figure~3: 
spherical planets with a band, a spot, obscuring rings (without ring shadows), a day-night difference, and a 
flattened (ellipsoidal) planet with a horizontally homogeneous atmosphere. 
Obviously, the parameter space for these calculations is enormous.
Here, we present a limited number of cases to provide an indication of the range of 
polarization signals of inhomogeneous planets and to draw some qualitative conclusions.
We model our horizontally inhomogeneous planets using two different atmospheric 
profiles, and hence two different one-dimensional radiative transfer calculations
(at several emission angles). Depending on the case (band, spot, etc.) and viewing
geometry, we assign one of the two profiles to the grid points on the planet.
We use a temperature gradient of 300 K/$\ln{p}$ everywhere on the planet, but 
 in the clear parts of the planet, temperatures are 250~K lower than in the cloudy
parts, which is a modest temperature contrast for hot Jupiters \citep{sho09}. In best-fit model calculations of the directly imaged planets around HR 8799 and 2M1207 by \citet{bar11a,bar11b} the temperature gradient at pressure higher than 0.1 bar is on average steeper than the 300 K/$\ln{p}$ used here, whereas below this pressure the temperature gradient is more shallow. The temperature profile shown in \citet{mar11} shows very similar behaviour for $T_\mathrm{eff}$=1000 K, $g$=30 m s$^{-2}$ and $f_\mathrm{sed}$=2, although temperature gradients are slightly less steep around 0.1 bar. Also best-fit atmospheres of transiting exoplanets have temperatures gradients around 300 K/$\ln{p}$ in the lower atmosphere \citep{mad09,mad10}.  In the near-infrared, most of the thermal emission originates from the region in the atmosphere between 0.1-1 bar for positive temperature gradients \citep[e.g.][]{sho09,mad10}, and hence our assumed temperature gradient does not seem unreasonably steep.
For the cloudy part of the atmosphere, a Rayleigh scattering cloud layer with 
unity scattering optical thickness is placed at the top of the atmosphere. 

Our calculations of ellipsoidal planets can be compared to those by \citet{sen10}. 
For a homogeneous cloudy atmosphere and an inclination angle of 90$^\circ$, 
our $P$ at $\lambda_1$ as a function of oblateness compares well with the 
$I$-band $P$ of \citet{sen10} for e.g.~$T_\mathrm{eff}=1800$~K with 
$\log(g)=4.5$ and $f_\mathrm{sed}$=2 (see their Figure~1). 
 Like \citet{sen10}, we can reach values of $P$ of several percent for extreme 
oblatenesses at $\lambda_1$, and roughly twice that value at $\lambda_2$. Longer wavelengths will give only slightly lower $P$s, given identical temperatures and optical depths, as is shown in Fig.~\ref{fig.spec} (the cloud optical thickness is kept constant with wavelength). The figure also shows the effect of low clouds, which results in inverted polarization spectra. Also note the large difference in polarization spectra between the two cases, whereas the flux spectra are very similar.  Calculations with 1-$\mu$m Mg$_2$SiO$_4$ particles, with a cloud optical thickness of unity at 1~$\mu$m and wavelength-variations over the plotted wavelength range determined by Mie theory, result in very similar polarization spectra, scaled down by a factor of $\sim$5.

For the planet with the equatorial cloud band, we varied the band's latitudinal width
and the planet's inclination angle. A clear planet with a dusty band shows a maximum 
$P$ (0.5\% at $\lambda_1$ and 2\% at $\lambda_2$) for a 40$^\circ$-60$^\circ$ wide band.
For a given band width, $P$ decreases with decreasing inclination angle, to increase again slightly when inclination angles get larger and the band 
reaches the limb. When the planet is seen exactly pole-on, $P=0$ again, as expected.
For a dusty planet with a clear band, $P$ can reach 0.5\% at $\lambda_1$ 
and 4\% at $\lambda_2$, depending on the inclination angle.

Except for very special geometries, rings will usually obscure part of a planet
either directly or through shadowing. We simulated the presence of cold, optically 
thick rings for a range of inclination angles by modeling the obscured part of the disk with a 200~K blackbody (see Figure~3).
For a ring-planet radius ratio similar to that of Saturn, $P$ is at most 0.3\% at $\lambda_1$ 
and 0.8\% at $\lambda_2$ at a ring inclination angle of $\sim$45$^\circ$. If the rings extend further outwards, and hence cover a larger fraction of the planet's disk, the maximum value of $P$ increases
by a few tenths of a percent. Note that we only modelled the obscuration here. The rings themselves can scatter the thermal radiation from the planet and might also emit significant polarized radiation themselves if they are hot and optically thick. A three-dimensional model of the planet and rings would be needed to model the interplay between planet and rings.

All three cases above have the planet's spin axis as the axis of symmetry. As a result, 
the disk-integrated polarized flux $U$ equals zero and the angle of polarization
$\chi$ is either parallel or perpendicular to the spin axis, depending on the particles'
microphysical properties and the temperature profile. We now briefly discuss two cases
without this type of symmetry.

We simulated an equatorial cloudy hot spot by a dusty square of $20^\circ\times20^\circ$ 
(latitude $\times$ longitude) (see Figure~3). As the planet rotates, $F$ varies mildly, while 
$P$ and $\chi$ vary significantly (see Figure~4). Here, $P$ reaches 0.1\% 
at $\lambda_1$, and 0.6\% at $\lambda_2$. 

The planet with the day-night difference has a dusty and a clear hemisphere (split along 
longitude lines, parallel to the terminator). As the planet rotates, $P$ depends  on which parts of the hemispheres are in view. The maximum $P$ equals
0.6\% at $\lambda_1$ and 4\% at $\lambda_2$, and $P=0$ when only one hemisphere is in view. $\chi$ varies similarly to the spot case (Figure 4).
Again, flux $F$ varies only mildy compared to $P$.

\section{Conclusions and opportunities for exoplanet characterization}

We have calculated polarization signals of thermal radiation that is scattered
by particles in exoplanet atmospheres. Important parameters that determine the degree of
linear polarization $P$ are the particles' polarization phase function, the optical
thickness and the altitude of the particles, and the temperature and its gradient.

For spatially resolved planets, $P$ is usually highest near the planet's limb. 
For spatially unresolved planets, inhomogeneities on a planet's disk can 
cause a net polarization signal. In the cases we have considered (i.e.~a band, a spot, 
obscuring rings, a day-night difference and a horizontally homogeneous flattened disk), $P$ was
typically above 0.1\%, and values of several percent were reached in favorable cases.
Combining different effects, e.g.~a band on a flattened planet, could increase $P$
even further. With a high cloud and a positive temperature gradient, $P$ is higher 
in gaseous absorption bands than in the surrounding continuum. Unfortunately, little flux is
emitted in those bands in this case. Together with telluric absorption, this will make the water absorption bands of planets with a positive temperature gradient less suitable for detection of infrared polarization in exoplanets using ground-based telescopes. Fortunately, $P$ is at most a factor of a few lower in the atmospheric windows, whereas the flux can be several magnitudes higher, giving more opportunities for detection there.

Currently, exoplanetary thermal radiation is detected either during secondary 
eclipses of transiting planets or through direct imaging of planets at large orbital 
distances. In the former case the combined light of the star and planet is measured, and 
although these usually tidally locked planets are expected to display large temperature
inhomogoneities, measuring a 1\% polarization signal of the planet will be very challenging, 
as sensitivities of $\sim$10$^{-6}$ need to be reached. Indeed, direct imaging, where the planet is spatially resolved from its star, promises
to be a more suitable method to detect polarized exoplanets, especially with new instruments like Gemini/GPI and VLT/SPHERE.

There are several reasons why planets like those in the HR 8799 system might be more prone 
to producing polarized signals than brown dwarfs, which can be polarized by a few percent. 
Firstly, the surface gravity of planets is lower than that of brown dwarfs, making them 
more flattened for a given rotation rate \citep[see also][]{mar11}. Secondly, broadband flux measurements suggest 
the HR 8799 planets to be more cloudy than most brown dwarfs, which might be a common feature of 
young planets \citep{mad11}. Thirdly, the effective temperatures in the planets'  atmospheres are lower than those in known polarized brown dwarfs and, given a certain 
temperature gradient, lower temperatures will yield higher $P$.

A detection of an infrared polarized signal will immediately confirm the presence of scattering 
particles in the planetary atmosphere. In addition, the polarization angle $\chi$ will reveal 
the components of the planet's spin axis normal to the line-of-sight for a zonally symmetric 
planet. For such planets, both $P$ and $F$ will show little variability. 
On the other hand, if the polarization is caused by moving clouds or hot spots, 
both $P$ and $\chi$ will vary in time, and stronger than $F$. Hence, a time series
of $P$ and $\chi$ will give insight into the underlying source, and a periodic signal
could even yield atmospheric rotation rates. However, some cases, like a banded and an oblate planet, may give very similar flux and polarization spectra and disentangling the two cases might be very difficult. Knowledge of the gravity on a planet will perhaps help in such a case, as the gravity is strongly connected to the possible oblateness.

Together with flux measurements, polarimetry can also constrain particle properties, 
like their scattering albedo and size. Using polarimetry, one could for instance  distinguish between absorbing iron particles and scattering silicate particles, which yield similar fits to flux spectra \citep{mad11}. Fig.~\ref{fig.spec} also shows that the polarization spectrum is much more sensitive to e.g.~cloud top heights than the flux spectrum. Furthermore, the broadband 
polarization is sensitive to the atmospheric temperature gradient, which is not easily 
obtained from broadband flux measurements. 

\acknowledgements
We thank Wiel Wauben making his scattering code available to us. We thank the anonymous referee for useful suggestions. We acknowledge financial support by the Netherlands Organisation for Scientific Research (NWO).


\clearpage

\begin{figure}[htp]
\centering
\resizebox{\hsize}{!}{\includegraphics{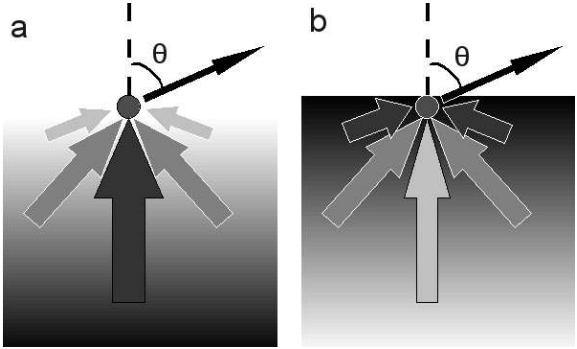}}
\caption{Sketch of thermal radiation with emission angle $\theta$, that is scattered  
         by a particle high in the planetary atmosphere, with dark areas          indicating higher temperatures and thus larger thermal fluxes. Panel (a)          illustrates a positive temperature gradient, and panel (b) a negative one.}
\label{fig.dia}
\end{figure}

\begin{figure}[htp]
\centering
\resizebox{\hsize}{!}{\includegraphics{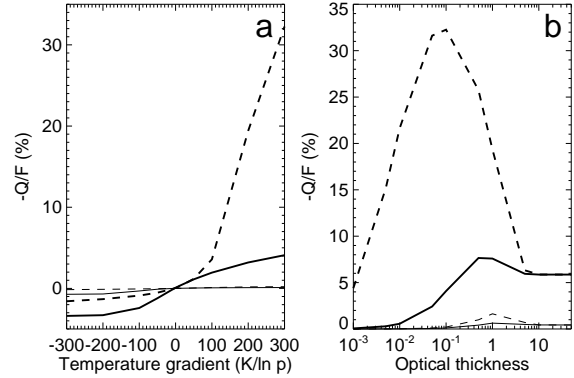}}
\caption{Normalized polarized fluxes $-Q/F$ at $\lambda=1.05 \mu$m (solid lines) and
         1.11 $\mu$m (dashed lines) for atmospheres with (a) different temperature gradients 
         and a high altitude cloud with optical thickness 0.1, and (b) a temperature gradient 
         of 300 $K/\ln{p}$ and different cloud optical thicknesses. Thick lines indicate 
         emission angles of 80$^\circ$ and thin lines those of 30$^\circ$.} 
\label{fig.results}
\end{figure}

\begin{figure}[htp]
\centering
\resizebox{\hsize}{!}{\includegraphics{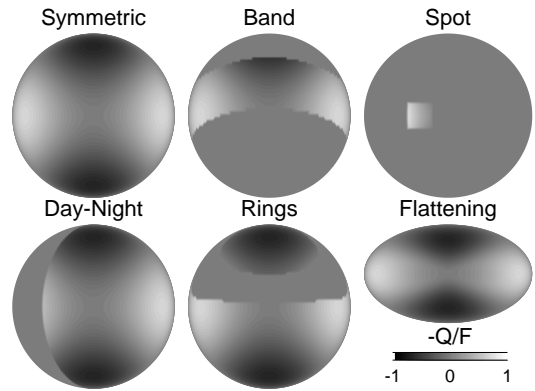}}
\caption{Simulated images of normalized $-Q/F$ across the planetary disk for the model planets presented 
         in this paper. Integrating $-Q/F$ over the disk yields $P=0$ for the spherically 
         symmetric planet, while it will usually leave a net $P$ for the other cases.}
\label{fig.examples}
\end{figure}

\begin{figure}[htp]
\centering
\resizebox{\hsize}{!}{\includegraphics{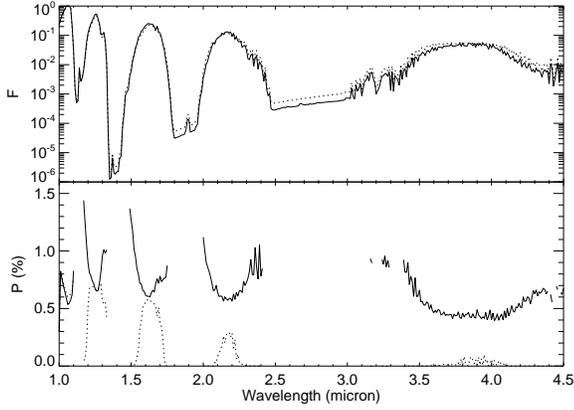}}
\caption{Normalized $F$ and $P$ as a function of wavelength for a homogeneous, flattened planet with oblateness of 0.3. Solid lines indicate models with clouds at the top of the atmosphere, dotted lines with clouds at 0.015 bar. $P$ is only plotted for $F>4 \cdot 10^{-3}$ as small integration errors at these low flux levels give rise to very noisy $P$ spectra there.} 
\label{fig.spec}
\end{figure}

\begin{figure}[htp]
\centering
\resizebox{\hsize}{!}{\includegraphics{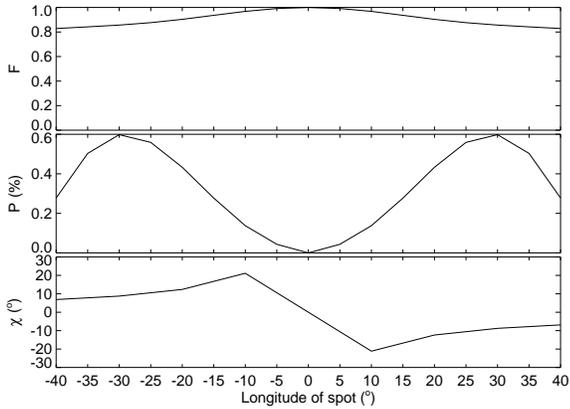}}
\caption{Normalized $F$, $P$, and $\chi$ at $\lambda_2$ of a rotating planet with
         a 20$^\circ$x20$^\circ$ dusty hot spot on its equator. The longitude denotes the distance of the spot with respect to the sub-observer point.} 
\label{fig.spot}
\end{figure}

\end{document}